\documentclass[sigconf,nonacm]{acmart}

\usepackage{color}
\usepackage{enumitem}

\AtBeginDocument{%
  \providecommand\BibTeX{{%
    \normalfont B\kern-0.5em{\scshape i\kern-0.25em b}\kern-0.8em\TeX}}}

\begin{document}

\title{LLM Comparator: Visual Analytics for Side-by-Side Evaluation of Large Language Models}

\newcommand{\name}[1]{\textit{LLM Comparator}}

\newcommand{\todo}[1]{\textcolor{orange}{[#1]}}

\definecolor{responseAColor}{rgb}{0.38, 0.51, 0.76}
\definecolor{responseBColor}{rgb}{0.89, 0.65, 0.24}
\definecolor{AIsBetter}{rgb}{0.35, 0.76, 0.95}
\definecolor{BIsBetter}{rgb}{0.99, 0.50, 0.50}
\definecolor{overlapping}{rgb}{0.38, 0.70, 0.38}


\settopmatter{authorsperrow=4, printfolios=true}

\author{Minsuk Kahng}
\email{kahng@google.com}
\affiliation{%
  \institution{Google Research}
  \city{Atlanta}
  \state{GA}
  \country{USA}
}

\author{Ian Tenney}
\email{iftenney@google.com}
\affiliation{%
  \institution{Google Research}
  \city{Seattle}
  \state{WA}
  \country{USA}
}

\author{Mahima Pushkarna}
\email{mahimap@google.com}
\affiliation{%
  \institution{Google Research}
  \city{Cambridge}
  \state{MA}
  \country{USA}
}

\author{Michael Xieyang Liu}
\email{lxieyang@google.com}
\affiliation{%
  \institution{Google Research}
  \city{Pittsburgh}
  \state{PA}
  \country{USA}
}

\author{James Wexler}
\email{jwexler@google.com}
\affiliation{%
  \institution{Google Research}
  \city{Cambridge}
  \state{MA}
  \country{USA}
}

\author{Emily Reif}
\email{ereif@google.com}
\affiliation{%
  \institution{Google Research}
  \city{Seattle}
  \state{WA}
  \country{USA}
}

\author{Krystal Kallarackal}
\email{kallarackal@google.com}
\affiliation{%
  \institution{Google Research}
  \city{Cambridge}
  \state{MA}
  \country{USA}
}

\author{Minsuk Chang}
\email{minsukchang@google.com}
\affiliation{%
  \institution{Google Research}
  \city{Seattle}
  \state{WA}
  \country{USA}
}

\author{Michael Terry}
\email{michaelterry@google.com}
\affiliation{%
  \institution{Google Research}
  \city{Cambridge}
  \state{MA}
  \country{USA}
}

\author{Lucas Dixon}
\email{ldixon@google.com}
\affiliation{%
  \institution{Google Research}
  \city{Paris}
  \country{France}
}

\renewcommand{\shortauthors}{Kahng, et al.}

\begin{abstract}
  Automatic side-by-side evaluation has emerged as a promising approach to evaluating the quality of responses from large language models (LLMs).
However, analyzing the results from this evaluation approach raises scalability and interpretability challenges.
In this paper, we present \name{}, a novel visual analytics tool for interactively analyzing results from automatic side-by-side evaluation.
The tool supports interactive workflows for users to understand when and why a model performs better or worse than a baseline model, and how the responses from two models are qualitatively different.
We iteratively designed and developed the tool by closely working with researchers and engineers at a large technology company.
This paper details the user challenges we identified, the design and development of the tool, and an observational study with participants who regularly evaluate their models.

\end{abstract}




\begin{teaserfigure}
  \includegraphics[width=\textwidth]{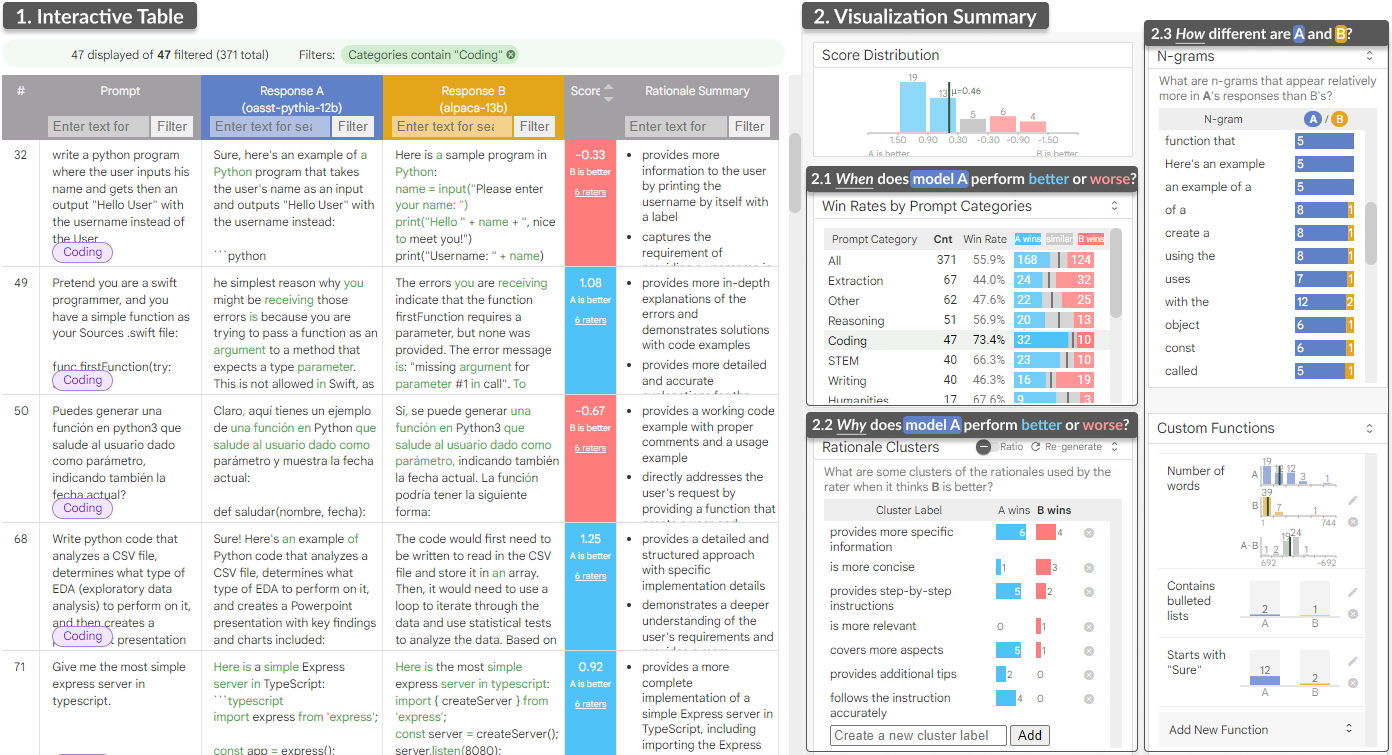}
  \vspace{-0.25in}
  \caption{\name{} enables model developers and researchers to interactively analyze results from \textit{automatic side-by-side evaluation} of large language models (LLMs). To evaluate the quality of responses from an LLM (A), users can compare them with those from a baseline LLM (B). The tool's \textit{interactive table (1)} enables users to inspect individual prompts and their responses in details, and its \textit{visualization summary (2)} supports analytical workflows that help understand \textit{when (2-1)} and \textit{why (2-2)} a model performs better or worse and \textit{how (2-3)} the two models' responses are different.}
  \label{fig:teaser}
\end{teaserfigure}


\maketitle

\begin{teaserfigure}
  \includegraphics[width=\textwidth]{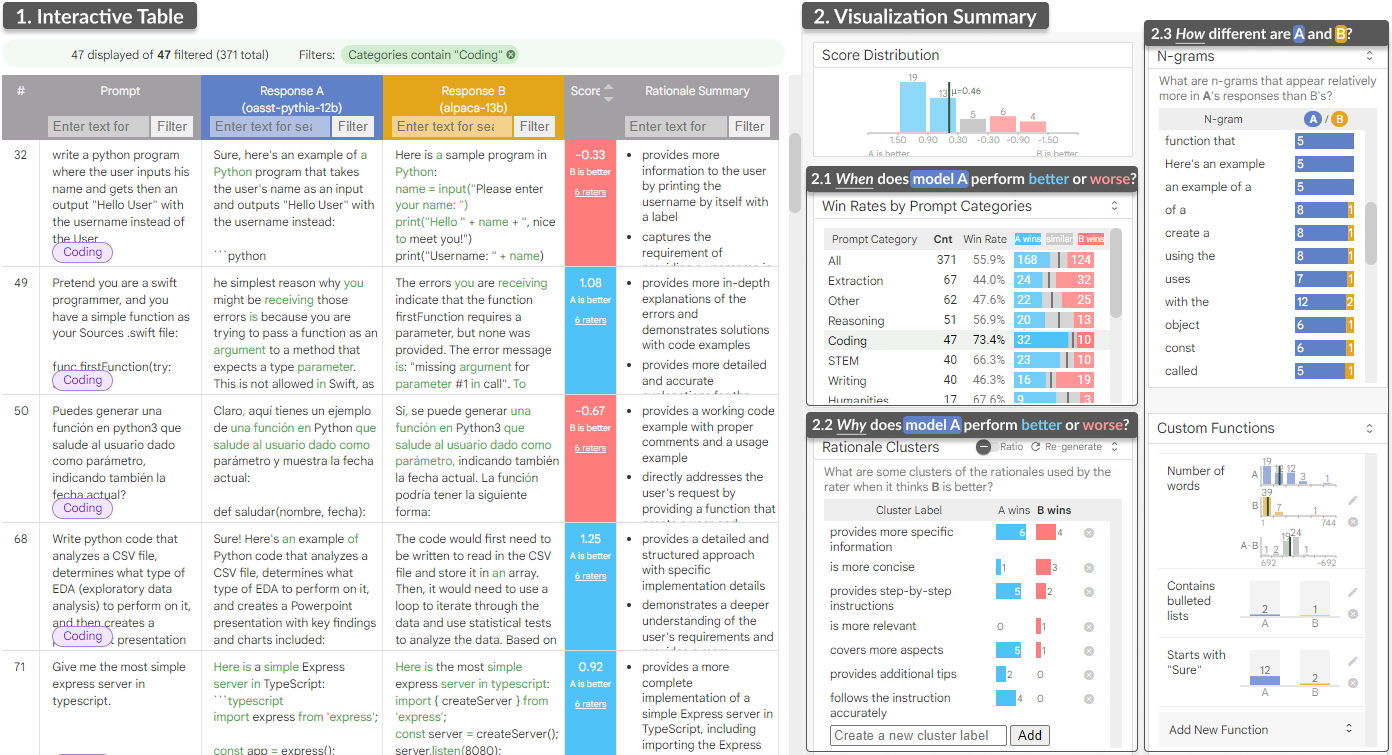}
  \vspace{-0.25in}
  \caption{\name{} enables model developers and researchers to interactively analyze results from \textit{automatic side-by-side evaluation} of large language models (LLMs). To evaluate the quality of responses from an LLM (A), users can compare them with those from a baseline LLM (B). The tool's \textit{interactive table (1)} enables users to inspect individual prompts and their responses in details, and its \textit{visualization summary (2)} supports analytical workflows that help understand \textit{when (2-1)} and \textit{why (2-2)} a model performs better or worse and \textit{how (2-3)} the two models' responses are different.}
  \label{fig:teaser}
\end{teaserfigure}

\section{Introduction}

Large language models (LLMs) are constantly being trained and tuned by researchers and engineers to improve their models. This involves adjusting model parameters, adding or removing training data examples, and other changes to the training procedure. A critical challenge for them is evaluating whether an updated model performs sufficiently well to supplant a baseline model.

However, the evaluation of LLMs pose unique challenges.
Unlike traditional machine learning models that are evaluated based on comparison with ground-truth answers, it is impractical to set ground truth responses for LLMs which generate long freeform text.
Therefore, a widely-used approach to evaluating LLMs is to ask humans to rate the output from the model by comparing with that from a baseline model. While effective, it does not scale to many experiments, as it is expensive to obtain human ratings.
To mitigate these challenges, \textit{automatic side-by-side evaluation (a.k.a., AutoSxS, LLM-as-a-judge)} has emerged as a promising approach to evaluating LLMs~\cite{vertex-autosxs,zheng2023judging}.
This approach involves asking another LLM to compare the quality of the outputs of two models. The prompt typically asks the LLM to select which response is better in terms of their quality.
Additionally, the LLM might be asked to justify its selection.

To deeply understand practitioners' workflows that utilize automatic side-by-side evaluation, we have had conversations with researchers and engineers in a variety of teams at a large technology company.
We learned that 
while aggregated scores 
from these LLM-based automatic raters 
quickly offer an initial assessment of model performance as a single number,
people have strong needs for further analysis about the rater results. They particularly raise interpretability and sensemaking challenges. For example, they want to understand \textit{why} a particular model received a score of 56\% of win rate. It is also difficult to deduce the types of scenarios in which a model will perform well or poorly.

In this paper, we present \name{}, a novel interactive tool 
for researchers and engineers
to analyze automatic side-by-side evaluation results in a scalable manner.
It provides interactive, visual analytics workflows for users to obtain the visual summary of the results from the side-by-side ratings while inspecting corresponding individual examples to explore the qualitative behavioral differences between the two models.
Specifically, the tool visualizes slice-level performances (\textbf{when} the model performs better), rationale summaries (\textbf{why} it is better), and n-grams and custom functions (\textbf{how} they are different).

\name{} has been successfully integrated into evaluation pipelines for large teams at Google. 
Since our initial announcement to select teams, the tool has attracted over 400 users within the first three months, facilitating the analysis of over 1,000 distinct automatic side-by-side experiments.
Their feedback has enabled us to iteratively improve the tool.
Section \ref{sec:tool} describes our latest prototype, and Section \ref{sec:study} presents a qualitative user study that evaluates it with six participants who regularly use automatic raters for their model developments.

\section{Current Workflows \& Design Goals}
\label{sec:challenges}

In this section, we discuss the current practice of LLM evaluations and our design goals for building a new interactive tool for analyzing the automatic side-by-side evaluations of LLMs. We base our discussion on our informal conversations with over 20 people from multiple teams at a large technology company
over the past year.

\begin{figure*}[!tb]
  \centering
  \includegraphics[width=1.0\linewidth]{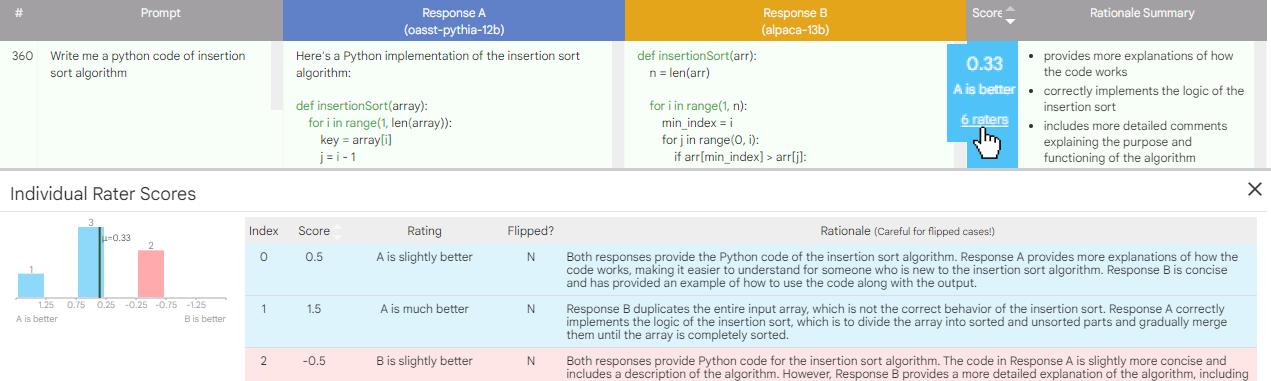}
  \vspace{-0.2in}
  \caption{Users can inspect the individual ratings to see the detailed rationales used by the raters.}
  \Description{}
  \label{fig:ratings}
\end{figure*}

\subsection{Current Practice of LLM Evaluations}

We discovered that automatic side-by-side evaluation (i.e., AutoSxS~\cite{vertex-autosxs}, LLM-as-a-judge~\cite{zheng2023judging}) is one of the most prevalent evaluation practices.
Once model developers train a new model, they would like to quickly evaluate it by running automatic side-by-side evaluation (AutoSxS) libraries, before conducting more costly and time-consuming human evaluations.
Specifically, the process consists of:

\begin{itemize}
\item \textbf{Baseline models:} When running AutoSxS libraries, the baseline model is set as a currently-deployed version of the model or one that has been shown to perform well (e.g., PaLM 2~\cite{anil2023palm}).
\item \textbf{Prompt sets:} People select one of the available prompt sets, each typically ranging from a few hundreds to thousands. These prompts are commonly tagged with slice categories (e.g., email writing, coding).
\item \textbf{Individual ratings:} The AutoSxS libraries take an input prompt, a response from the model to test, and a response from the baseline. It returns a Likert-scale rating (e.g., ``A is much better,'' ``B is slightly better,'' etc.) accompanied by a brief rationale. The Likert-scale ratings are transformed into numeric scores (e.g., ``A is much better'' equals 1.5, ``A is better'' equals 1.0, ``B is much better'' equals -1.5). To counteract the stochastic nature of the scores, the process is often repeated multiple times, and the final score is calculated as the average of these repetitions~\cite{liu2024lipo,zheng2023judging}. 
\item \textbf{Aggregated metrics:} The libraries calculate metrics from the ratings across a set of prompts, with average scores and win rates being the most prevalent. A win rate can be defined as the proportion of scaled rating scores that are above or below the threshold (A wins if score > 0.3; B wins if < -0.3; tie otherwise).
\end{itemize}

\subsection{User Challenges in Analyzing Evaluation Results}

We have identified common workflows of how model developers and researchers analyze the results from these AutoSxS libraries:

\begin{itemize}[leftmargin=0.3in]
\item No specialized tools existed for the analysis of AutoSxS results. Typically, results are loaded into spreadsheets---with one row for each input prompt, and the columns are prompt, response A, response B, and the score. In addition, some practitioners load this data into \textit{computational notebooks} (e.g., Jupyter, Colab).
\item 
Practitioners eyeball individual examples (i.e., prompts and LLM responses) to interpret evaluation results and compare differences between responses from two models qualitatively.
They either randomly select examples to examine or
sort them by scores using spreadsheets and examine those with exceptionally high or low scores. 
However, it is often challenging to read and compare these texts, as spreadsheets are not designed for long texts.
\item They have strong interests in computing metrics (e.g., average scores, win rates) by slices (e.g., prompt categories) to identify which slices underperform or outperform relative to others.
Then they often want to inspect examples in these slices, but it requires switching between different tools.
\item For further analysis, practitioners compute additional features from response texts (e.g., number of tokens) and aggregate the feature values across the examples by using computational notebooks or other tools.
\end{itemize}

Importantly, both detailed examination of individual examples and analysis of aggregated data are essential; however existing tools fail to connect these two types of analyses.

\subsection{Design Goals}

Given the current practice of model evaluations and the challenges in analyzing the evaluation results outlined above, we distilled the following design goals for building tools for side-by-side evaluation analysis:

\begin{enumerate}[label=DG\arabic*.,leftmargin=0.3in]
\item Facilitate interactions between the aggregated information and individual examples. This will enable users to identify their slices of interests in diverse ways and examine specific prompts and their response pairs in details.

\item Provide workflows to uncover answers to the following analytical questions:

\begin{enumerate}[label=2-\arabic*.]
\item When: In which situations does model A perform better than model B?
\item Why: What are the common rationales used by the raters? Why does it say one model is better than another?
\item How: How are the responses between two models different?
What qualitative patterns can be observed? Can these patterns be used to inform data or model improvements?
\end{enumerate}

\item Perform the analysis of evaluation results at scale for a large number of prompts. This will allow users to more confidently discern the performance differences between two models.

\end{enumerate}

\section{Visualization Design and Development}
\label{sec:tool}

In this section, we introduce \name{}, a interactive tool for the side-by-side comparison of LLMs. 
\autoref{fig:teaser} shows our tool's interface for a scenario where a researcher who develops a new LLM evaluates its performance by comparing it to a baseline model. For a prompt set obtained from public benchmarks~\cite{zheng2023judging}, they obtain automatic side-by-side evaluation ratings using a third LLM\footnote{We used Google Cloud's Generative AI APIs on Vertex AI available at \url{https://cloud.google.com/vertex-ai/docs/generative-ai/learn/overview}.} to compare the quality of response pairs.
The tool consists of two main panels: the (1) an \textit{interactive table} for detailed individual example inspection and (2) a \textit{visualization summary} for overviews and filtering options that support the users' analytical workflows.

\subsection{Interactive Table}

Each row in the table represents a prompt, its corresponding responses from two models, the rater's score, and a rationale summary. Below we highlight a few unique features of the interactive table:

\begin{itemize}[leftmargin=0.3in]
\item \textbf{Overlapping word highlights.} To facilitate quick and easy comparison of two response texts, we highlight overlapping words between the two
as {\color{overlapping}green text} (e.g., ``def insertionSort'' in Figure \ref{fig:ratings}).
\item \textbf{Rationale summary.} The rationale is typically too lengthy to read in full, particularly with multiple raters involved (shown in Figure \ref{fig:ratings}, bottom). To address this challenge, we employ another LLM to summarize the rationales into a bulleted list (in Figure \ref{fig:ratings}, rightmost column). If a row receives six ratings and the average outcome favors A (with 4 for A being better and 2 for B), we ask the LLM to summarize the four cases favoring A.
\item \textbf{Option to see the detailed rating results.} The average score is displayed on the table row, with an option to view detailed results if desired (i.e., by clicking ``6 raters'' link as shown in \autoref{fig:ratings}).
\item \textbf{Color coding scheme.} We represent A with {\color{responseAColor}indigo} and B with {\color{responseBColor}orange}. Also, to represent the rater's decisions,
we use {\color{AIsBetter}blue} to indicate rows where the rater prefers A, {\color{BIsBetter}red} where the rater prefers B, and {\color{gray}gray} to denote ties.
\end{itemize}

\subsection{Visualization Summary}

The visualization summary panel features several components designed to support the analytical workflows of users:

\textbf{Score Distribution.}
Upon first encountering the summary metric (e.g., average score $=$ 0.46), users would often ask about its detailed distribution. To help answer this question, we display a simple histogram for the distribution of scores from the automatic raters (ranging from 1.5 to -1.5).

\textbf{Win Rates by Prompt Category (\textit{when}).}
To answer the common analysis question of in what scenarios model A outperforms or underperforms compared to model B (DG2-1), we present a visualization of performance across different prompt categories.
This helps users to identify prompt categories with notably higher or lower scores, informing which data examples to inspect further.
In Figure 1 (\textbf{2-1} on the right), a high-scoring category ``Coding'' is selected.

\begin{figure}[!tb]
  \includegraphics[width=0.75\columnwidth]{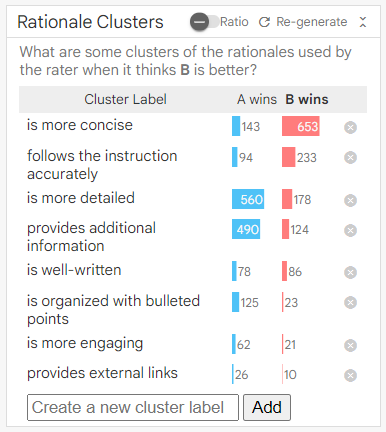}
  \caption{The rationale clusters view presents a list of rationales that are frequently used by the automatic rater. Users can dynamically add ones to compare the occurrences of relevant rationales between the two models.}
  \Description{}
  \label{fig:rationale-clusters}
\end{figure}

\textbf{Rationale Clusters (\textit{why}).}
To help users understand the rationales behind the rater's decisions (DG2-2), we condense a large array of rationales into several representative themes.
While there are various methods to produce these themes, for example, by running clustering on all rationale bullets and subsequently labeling the clusters, we opted for a novel LLM-based approach that performs better and runs faster based on our testing. Specifically, we first ask a different LLM to generate a set of diverse and representative cluster labels given a sample of the rationale bullets, inspired by recent work~\cite{zhong2022describing, wang2023goal}.
We then assign each rationale bullet to clusters (represented by their labels) based on embedding similarity,\footnote{To assign rationale bullets into clusters shown in Figure \ref{fig:teaser}, we used Google Cloud's text-embeddings APIs at \url{https://cloud.google.com/vertex-ai/docs/generative-ai/embeddings/get-text-embeddings}.} i.e., if the cosine similarity between the bullet and the label exceeds an empirically determined threshold, it is considered a match. Note that each bullet can be assigned to multiple clusters.

As shown in~\autoref{fig:rationale-clusters},
for each cluster label, the tool counts the number of instances where model A is determined to be better, and vice versa. By sorting these counts, users will be able to identify common rationales used by the rater.
Moreover, it can also be particularly useful to examine the ratio between the count for A and B. 
For instance, if a certain rationale cluster (e.g., ``is more concise'') shows a significantly higher count for B, users can hypothesize that B's responses are generally more concise than A.

Users can interact with the clustering results by dynamically adding or removing individual clusters, or regenerating the entire cluster set for only the filtered examples.
In addition, this analysis can be combined with the prompt category filter, enabling users to inspect the different rationales used by the rater for different types of prompts.

\textbf{N-grams and Custom Functions (\textit{how}).}
To grasp the nuances of the rationales, users must be able to examine individual instances.
For instance, if the automatic rater states that ``B is more organized,'' a user might still wonder about what it means to be ``organized.''
While users can directly inspect individual response pairs from the table, we provide additional support for such analysis with \textbf{n-gram analysis} and \textbf{custom functions}.

\begin{figure}[!tb]
  \includegraphics[width=1.0\columnwidth]{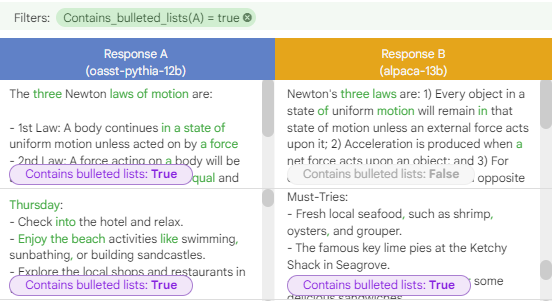}
  \caption{Users can dynamically create functions that apply to responses. In this example, a function specified as a regular expression (i.e., \texttt{"\textbackslash n([*-])\textbackslash s"}) checks if each response contains bulleted lists, whose results are displayed as purple chips.}
  \Description{}
  \label{fig:chips}
\end{figure}

\begin{itemize}[leftmargin=0.3in]
\item 
\textbf{N-gram Analysis.} The tool presents frequently occurring n-grams (n$=$1 to 7) in responses from either A or B, compared to its counterpart (e.g., ``Here's an example'' appears 150 times in A's responses while appearing only 3 times in B's).

\item
\textbf{Custom Functions.} Alternatively, users can define \textit{custom functions} represented as regular expressions (e.g., newline character followed by dash or star indicating bulleted items) or JavaScript function expressions (e.g., word count specified by ``\texttt{output.split(/\textbackslash s+/).length}'').
Upon specifying these expressions, they immediately apply to each individual response and return either boolean or numeric values. For boolean values (e.g., whether it contains bulleted items), the tool visualizes the results as percentage bar charts; for numeric values (e.g., word count), it displays histograms. They can be displayed on top of the responses when selected (as shown in Figure \ref{fig:chips}).
\end{itemize}

\subsection{Implementation}

\name{} is implemented as a web-based application.
Its preprocessing module loads a data file that stores the results from the AutoSxS libraries containing a list of prompts with response pairs and the ratings with rationales.
Then it calls an LLM to summarize rationales into bullet points, generate cluster labels, and compute embeddings to be used for cluster assignments.
The server, written in Python, loads this preprocessed data file and then transmits it to the client in JSON format. Once data is loaded into the client, all computations, such as filtering, sorting, and cluster assignments, are performed dynamically on web browser. The client-side code is written in TypeScript using the Lit webcomponents framework.\footnote{\url{https://lit.dev}} When a user requests to regenerate rationale clusters, the server invokes calls to an LLM using a RPC call.

\subsection{System Deployment}
\name{} has been developed based on iterative feedback from many engineers, researchers, and data scientists at Google.
Since our initial announcement to select internal LLM development teams, the tool has attracted over 400 users, facilitating the analysis of over 1,000 distinct side-by-side evaluation experiments.
In addition, it has been deployed on evaluation pipelines for large teams who develop LLMs for their products.
The final-stage of the pipelines
performs preprocessing for \name{}. When the pipeline is complete, users see a direct link to our tool on the platform interface.

While the earlier versions of the tool featured the interactive table and a subset of the visualization summary components described in this section,
the latest prototype updated based on the user feedback offers the full functionalities described in this section, including the rationale clusters and N-grams analysis.
In the next section, we present our evaluation of this latest prototype.

\section{Observational Study}
\label{sec:study}

We conducted an observational study to investigate how users would use \name{} presented in Section \ref{sec:tool}.

\subsection{Study Setup}

\indent\indent
\textbf{Participants.}
We recruited six participants (P1-6) from our company, consisting of software engineers, researchers, and data scientists who are directly involved with teams dedicated to developing LLM-based products.
All participants had experience conducting automatic side-by-side evaluations within the past month.
Additionally, some had previous experience with earlier iterations of our tool, e.g., versions without the  rationale clusters feature.

\textbf{Study Protocol.}
Each study session was conducted remotely over video conferencing and took around 45 minutes.
After participants sign the consent form,
we first conducted a 10-minute interview focused on the participant's role in model evaluation.
This was followed by a 5 to 10-minute tutorial on the tool's features.
Participants were then asked to use the tool while think aloud to analyze a recent side-by-side evaluation run they had performed on internal evaluation platforms.
The session concluded with a short reflective interview, and 
the participants were compensated with \$25 USD.
We analyzed the results through thematic analysis.

\subsection{Key Usage Patterns}

Our study revealed the following interesting usage patterns.\footnote{To honor participants' requests for confidentiality, we have redacted certain details about the models, data, and prompt categories. Despite this, the general patterns of use remain accurately represented.}

\subsubsection{\textbf{Example-first deep dive.}}

P1 and P2 invested significant time in reading prompts and responses to gain insights from the results, especially when they first launched the tool.
Driven by the overall metric favoring Model B (baseline model), P2 wanted to inspect low-scoring examples for Model A (their model).
They sorted the examples by the rater score and scanned the rows one by one.
They meticulously read prompts to find one they can familiarize with, 
and then read and compared response pairs.
P2 said this process is crucial because the automatic rater is not always right, so they need to make sure if the rater is working correctly.
P1 used an interesting alternative strategy. They concealed the score column and predicted the automatic rater's scores, mimicking the process for human raters.

After spending time analyzing examples, participants began formulating hypotheses about behavioral differences between the two models.
P2 noticed a case where Model A's response succinctly only include the code for a prompt about coding, while B additionally provided detailed explanations.
This difference caught P2's attention, because it might be caused by a specific change they have made to their model.
To further find similar cases,
they filtered examples by prompt category (i.e., coding) and quickly found several other examples that exhibit similar patterns.
Moreover, the rationale clusters view reveals one named ``Provide clear explanations'' with much higher counts for B, further confirming their hypothesis.

\subsubsection{\textbf{Prior experience-based testing.}}

P3, P4, and P5 leveraged their prior knowledge to identify undesirable model behaviors.
P3 sought to find responses containing phrases like ``I'm sorry'' or ``unfortunately'' which often signal a model's refusal to answer the tasks specified in prompts. They need to discern whether these are genuinely unanswerable prompts for LLMs or areas for potential improvement to deliver accurate responses.
Similarly, P5 noted the desire to detect cases
where LLMs generate unnecessary phrases (e.g., starting sentences with ``here is'', overusing bold text) to internally optimize their objectives, which is a known behavior of LLMs~\cite{amodei2016concrete}.

Participants reported maintaining a collection of such undesirable patterns for testing purposes (similar to performing \textit{testing} in software engineering~\cite{zhang2020machine}), and used the tool to determine if these patterns were present in either side of the models.
Specifically, they used the tool's N-grams and custom function features to make initial searches for these phrases. Subsequently, they used visualizations to compare the occurrences of these phrases across the two models. For example, after finding a noticeable difference between counts, they meticulously examined the corresponding individual responses and leveraged the rationale summaries and clusters to check whether the automatic raters paid attention to this information.

\subsubsection{\textbf{Rationale-centric top-down exploration.}}

The rationale clusters view enabled a diverse set of ways to analyze data that were previously unavailable.
P2 had used the earlier versions of the tool before, and they used it primarily for inspecting individual examples. While the only way to understand the rater's rationales was selecting an example first and opening the detailed rating results view, the updated version introduces rationale clusters, providing new methods for in-depth data investigation to validate their hypotheses about the model's behavior.
In addition, P3, who had also used the tool many times before, first searched for specific keywords like ``sorry'' as described earlier. However, they later noticed one of the rationale clusters ``Avoids harmful content'', and by applying a filter for this cluster, they were pleased to see several interesting keywords from the N-grams components. These keywords include those which they had to manually search for individually, including ``I'm sorry.''

Participants also adopted a more exploratory approach actively engaging with the visualizations to discover interesting patterns.
Coordinated views that are dynamically updated capture their attention and spark curiosity.
For instance, P6 noticed a category with a significantly higher win rate from the chart. Applying a filter for this category, 
they could naturally form new hypotheses from one of the rationale clusters about conciseness. This led them to use a custom function for word count 
and identified responses that are very short and problematic.

\subsection{Discussions and Future Opportunities}

In addition to the above usage patterns, we uncovered opportunities to further enhance users' analysis workflows.

\textbf{LLM-based custom metrics.} While the N-grams and custom functions are effective for analyzing qualitative differences, people have additional needs to assess high-level attributes (e.g., safety, verbosity).
To address this limitation, we can employ yet another LLM,
similar to prior work~\cite{kim2023evallm}. However, this approach brings substantial complexity due to the extensive LLM calls, particularly for the dynamic evaluation of large prompt sets. Exploring practical solutions to mitigate this bottleneck would greatly enhance the feasibility and scalability of LLM-based evaluations.

\textbf{Pre-configured undesirable patterns.}
Participants expressed a strong desire for the tool to be pre-configured with specific unwanted patterns, to avoid manually defining new functions or searching for individual keywords.
For example, P3 particularly cared about identifying the issue of repeated sentences in LLM outputs, highlighting the importance to be able to easily detect and flag such occurrences.

\textbf{Improving rationale clustering.}
The pipeline for clustering rationales relies on LLM calls, which could be error-prone.
Also, it could be less than ideal to use embedding similarity for clustering assignments, as embeddings reflect not only \textit{semantic} but also \textit{syntactic} similarity.
Alternative computational approaches 
and more advanced interactions (in addition to what we implemented, e.g., adding new clusters) would boost the robustness and efficiency of this pipeline.

\section{Related Work}

\indent\indent
\textbf{Visual Analytics for Machine Learning Interpretability.}
In the past decade, a variety of methodologies and tools for machine learning analysis have been developed from the visualization community.
These include early works that emphasized the importance of visualizing individual data points~\cite{amershi2015modeltracker} and supporting slice-level analysis~\cite{kahng2017activis, ming2018rulematrix},
tools that utilized various interpretability methods to explain individual predictions~\cite{tenney2020language}, and methods and techniques for model comparison~\cite{gleicher2020boxer, wang2022learning, strobelt2021lmdiff, boggust2022embedding}.
As LLMs have emerge, tools targeting specific types of language models have been introduced~\cite{coscia2023knowledgevis, strobelt2021lmdiff, strobelt2022interactive, brath2023role}.

\textbf{Interactive Tools for LLM Evaluations.}
With ChatGPT's rise in 2023, interactive tools for LLM evaluations and comparisons have begun to appear in late 2023.
A recent preprint, ChainForge~\cite{arawjo2023chainforge}, presented a flexible framework to perform comparisons with user-specified functions.
Another recent work, EvalLM~\cite{kim2023evallm}, presented a tool for interactively performing LLM-based evaluations by user-defined criteria.
Different from these concurrently developed approaches, 
our work focuses on the visual analysis and interpretation of large-scale evaluations for industry practitioners.

\section{Conclusion}
We presented a new interactive tool for analyzing results from automatic side-by-side evaluation methods.
The tool aimed at enabling users to analyze when and why a model performs better or worse than a baseline model and how they behave differently.
Our observational study indicated that the tool enables participants to form hypotheses about the automatic ratings, verify known model behaviors, and analyze qualitative differences between model responses.

\begin{acks}
We thank Sujeevan Rajayogam, Fernanda Vi\'{e}gas, Martin Wattenberg, Timothy Chung, Ankur Taly, and our colleagues at Google's People + AI Research (PAIR) team for their support and feedback. We also thank users of \name{} for their feedback and suggestions.
\end{acks}

\bibliographystyle{ACM-Reference-Format}
\bibliography{references}


\end{document}